\documentclass[11pt,a4paper]{article}
\usepackage[T1]{fontenc}
\usepackage[utf8]{inputenc}
\usepackage[margin=25mm]{geometry}
\usepackage{lmodern}
\usepackage{microtype}
\usepackage{graphicx}
\usepackage{array,longtable,booktabs,ragged2e}
\usepackage{pdflscape}
\usepackage[numbers,sort&compress]{natbib}
\usepackage{xurl}
\usepackage[hidelinks]{hyperref}
\usepackage[font=small,labelfont=bf]{caption}
\hypersetup{pdftitle={When Is Content AI-Generated Enough? Labelling Synthetic Media under the Digital Services Act and the AI Act},pdfauthor={Marie-Therese Sekwenz}}
\title{When Is Content ``AI-Generated Enough''? Labelling Synthetic Media under the Digital Services Act and the AI Act}
\author{Marie-Therese Sekwenz\\[3pt]\small TU Delft, Delft, The Netherlands\\\small\texttt{m.t.sekwenz@tudelft.nl}}
\date{Preprint -- 7 September 2026}
\begin{document}
\maketitle
\begin{center}
\small Accepted and presented at the \href{https://2026.ecafconference.org/}{Fifth European Conference on Algorithmic Fairness (ECAF 2026)}, Ghent, Belgium, 2--4 September 2026.
\end{center}
\begin{abstract}
European platform and AI governance increasingly relies on transparency duties to address synthetic and manipulated media. Under the DSA, very large online platforms and search engines may use prominent markings and recipient-facing indication tools as systemic-risk mitigation measures. Under the AI Act, providers must support machine-readable marking, while deployers must disclose deepfakes and certain AI-generated or manipulated public-interest text, subject to statutory qualifications.

This extended abstract examines when labelling is a meaningful regulatory response to synthetic media and when it risks becoming over-inclusive, under-inclusive, or ineffective. It argues that the central challenge is not only whether content should be labelled, but how legal thresholds, technical provenance systems, platform interfaces, and reporting practices determine when content is sufficiently generated, manipulated, or authentic-looking to trigger transparency obligations. Drawing on the emerging Article~50 AI Act implementation framework and a snapshot of the DSA Statement of Reasons database, the paper identifies four governance tensions: definitional ambiguity, interface and responsibility design, communicative effectiveness, and fairness and contestability. It conceptualises labelling as a socio-technical classification practice that distributes responsibility among AI providers, deployers, platforms, uploaders, and recipients.
\end{abstract}

\medskip
{\small\textit{Scope note.} This extended abstract analyses the second draft Code of Practice on Transparency of AI-Generated Content (March 2026) and a DSA dashboard snapshot collected on 30 April 2026. References to proposed measures and forthcoming guidance are relative to that implementation stage.}

\section{Synthetic Media as a Regulatory Object}
Synthetic and manipulated media have become a central object of European digital regulation. The increasing accessibility of generative AI tools means that images, audio, video, and text can be produced or altered in ways that challenge ordinary expectations of authenticity \cite{hua_generative_2024}. Regulatory responses often turn to labelling: rather than prohibiting all synthetic media, the law requires certain actors to disclose artificial generation or manipulation \cite{noauthor_regulation_2022,noauthor_commission_2024,noauthor_regulation_2024}. Labelling therefore appears as a compromise between freedom of expression, artistic freedom, innovation, and protection against deception, manipulation, and disinformation \cite{jamieson_flagging_2025,devadas_experimental_2024,disalvo_social_2022}.
Yet labelling is not self-executing. It requires decisions about what counts as synthetic or manipulated content, who should label it, what the label should say, where it should appear, and how users are expected to interpret it. These questions are difficult because contemporary media practices rarely fit a binary distinction between authentic and fake. A video may contain AI-generated subtitles, automated filters, synthetic voice correction, edited backgrounds, AI-enhanced lighting, or fully generated persons. Similarly, a political image may be AI-generated, digitally edited, staged, taken out of context, or misleadingly framed. In such cases, the regulatory problem is not only deception by deepfakes, but the instability of the categories through which deception is governed.
This text asks: \textit{How do the DSA, the AI Act, and the emerging implementation framework around Article~50 AI Act conceptualise deepfake transparency, and what practical problems arise when legal labelling duties are translated into technical marking systems, platform interfaces, and user-facing disclosures?}
\section{Regulatory and Implementation Background}
The DSA and the AI Act approach synthetic and manipulated media from related but distinct angles. The DSA addresses deepfake-like content within the systemic risk mitigation duties of very large online platforms and very large online search engines. Article~35(1)(k) DSA identifies, among the measures available to mitigate systemic risks under Article~34 DSA, measures ensuring that generated or manipulated image, audio, or video content that appreciably resembles existing persons, objects, places, entities, or events and falsely appears authentic or truthful is distinguishable through prominent markings on online interfaces. Importantly, the same provision also requires ``an easy to use functionality'' enabling users to indicate such information \cite{noauthor_regulation_2022}. This connects synthetic-media governance to interface design: platforms must not only display labels, but also provide mechanisms through which users (also experts like trusted flaggers according to Art. 22 DSA) can signal that content may be generated or manipulated, similar to content reporting bearing the risk of burdening the user, jargon, or uncertainty \cite{sekwenz_it_2025, sekwenz_there_2026}.
The AI Act addresses synthetic media through transparency obligations for providers and deployers of AI systems. Article~3(60) AI Act defines a ``deep fake'' as AI-generated or manipulated image, audio, or video content that resembles existing persons, objects, places, entities, or events and would falsely appear to a person to be authentic or truthful \cite{noauthor_regulation_2024}. Article~50(4) AI Act requires deployers\footnote{Under Article~3(4) AI Act, a deployer is a natural or legal person, public authority, agency, or other body using an AI system under its authority, except where the AI system is used in the course of a personal non-professional activity.} of AI systems that generate or manipulate image, audio, or video content constituting a deep fake to disclose that the content has been artificially generated or manipulated. It also applies to AI-generated or manipulated text published with the purpose of informing the public on matters of public interest, subject to exceptions for human review, editorial control, and editorial responsibility.

The Code of Practice is divided into rules on marking and detection applicable to providers\footnote{For providers, the Code emphasises machine-readable marking, detection mechanisms, digitally signed metadata, watermarking, interoperability, reliability, robustness, and accessibility \cite[pp. 8–19]{CodePracticeTransparencyAIGeneratedContent2026SecondDraft}. } under Article 50(2) and (5), and rules on user-facing labelling applicable to deployers\footnote{For deployers, it focuses on user-facing disclosure through icons, labels, or disclaimers for deepfakes and certain AI-generated or manipulated text published with the purpose of informing the public on matters of public interest \cite[pp. 26–33]{CodePracticeTransparencyAIGeneratedContent2026SecondDraft}. } under Article 50(4) and (5) AI Act \cite[pp. 2, 5, 26]{CodePracticeTransparencyAIGeneratedContent2026SecondDraft}. This illustrates that labelling is not merely a visible warning attached to content, but part of a broader provenance and compliance infrastructure connecting AI-system providers, deployers, platforms, and end-users.
At the same time, the Code does not resolve all threshold questions. It states that elements related to the scope of key definitions and exceptions will be addressed in forthcoming Commission guidelines on Article~50 AI Act, developed in parallel \cite[pp.~4, 27]{CodePracticeTransparencyAIGeneratedContent2026SecondDraft}. This is analytically important because many of the most difficult labelling questions concern precisely these threshold issues: when content is sufficiently generated, manipulated, authentic-looking, or public-interest-oriented to trigger disclosure.
Building on this regulatory and implementation background, the paper makes three contributions. First, it conceptualises deepfake labelling as a socio-technical classification practice rather than merely a disclosure obligation. Second, it identifies four governance tensions arising from the
operationalisation of labelling duties: definitional ambiguity,
interface and responsibility design, communicative usefulness, and
fairness and contestability. Third, it uses the DSA Statement of Reasons database to examine how platforms report synthetic-media moderation in practice and to assess the relative role of labelling compared with removal, demotion, and other restrictions using a dashboard snapshot collected on 30 April 2026 (Appendix~\ref{screenshots}).
\section{Labelling as Provenance and Platform Governance}
The second draft Code of Practice makes visible a shift from labelling as a simple user-facing transparency measure toward labelling as a provenance infrastructure.
For example, the Code encourages richer provenance metadata identifying the AI system and provider,\footnote{On the provider side, the Code generally requires a multi-layered machine-readable marking approach consisting of digitally signed metadata and imperceptible watermarking, while permitting fingerprinting or logging as optional supplementary mechanisms \cite[pp. 8–10]{CodePracticeTransparencyAIGeneratedContent2026SecondDraft}.} the time of generation or manipulation, and, for manipulated content, the type of operation performed \cite[p. 11]{CodePracticeTransparencyAIGeneratedContent2026SecondDraft}.
On the deployer side, it proposes clear and distinguishable disclosure through icons, labels, or disclaimers, including a possible common EU icon and a second layer of more detailed information about what has been manipulated \cite{CodePracticeTransparencyAIGeneratedContent2026SecondDraft}.

The Commission's electoral guidelines under the DSA reinforce this broader view of labelling. They recommend that providers of VLOPs and VLOSEs whose services can be used to disseminate deceptive, false, or misleading generative AI content clearly label, or otherwise make distinguishable through prominent markings, synthetic or manipulated images, audio, or videos that falsely appear authentic or truthful. They further recommend standard and easy-to-use interfaces and tools for users to add labels to AI-generated content, efficient labels that are easily recognised by users, and measures ensuring that labelled content retains its label when shared by others on the same platform \cite[p.~16--17]{noauthor_commission_2024}.
Labelling is therefore linked to a broader set of mitigation measures, including marking, demotion, removal, detection through watermarks and metadata, and the adaptation of content moderation processes and algorithmic systems.
This implementation context reveals an important asymmetry. Current labelling structures often focus on the uploader or content creator.

Users who upload content may be asked to self-disclose whether the content was generated or modified with AI, for example through an upload toggle, disclosure field, or advertising-label option. Uploader-side labelling is important because it can make disclosure part of the ordinary publication workflow.
\section{EU Icon Categories and Regulatory Thresholds}
Table~\ref{tab:eu-icons-legal-framework} situates the emerging EU icon categories within the combined regulatory architecture of the AI Act, the Digital Services Act, and the Code of Practice on Transparency of AI-Generated Content. These instruments address synthetic and manipulated media from different but overlapping perspectives.

As Table~\ref{tab:eu-icons-legal-framework} shows, the implementation framework distinguishes between fully AI-generated content and pre-existing human-created content that has been partially modified using AI. The ``AI GENERATED'' icon is intended for content generated entirely through AI without human-created elements or human editorial control apart from prompting, while the ``AI MODIFIED'' icon applies where existing content has been partially altered using AI. The basic AI icon may be combined with a customised textual disclosure or an interactive second layer providing more detailed information about the nature and location of the AI involvement
\cite[pp.~33, 37]{CodePracticeTransparencyAIGeneratedContent2026SecondDraft}.
The categories nevertheless leave important threshold questions unresolved. Neither Article~50 AI Act nor Article~35(1)(k) DSA makes every use of AI subject to the same visible disclosure requirements. Minor enhancement, noise reduction, automated subtitling, colour correction, or other forms of AI-assisted editing may not turn content into a deepfake or qualifying manipulated public-interest text.

The distinction between full generation, material modification, and minor AI assistance is therefore central to determining when content is sufficiently ``AI-generated'' or ``AI-modified'' to trigger disclosure.
Table~\ref{tab:eu-icons-legal-framework} also illustrates that visible icons constitute only one layer of the regulatory framework. Machine-readable metadata, watermarking, provenance information, and detection mechanisms support technical identification, while platform interfaces determine how labels are displayed, how suspected synthetic media is reported, and how classification decisions may be contested.

\section{Fairness, Differential Effects, and Contestability}
Labelling also raises questions of algorithmic fairness that extend beyond whether a classification is technically accurate. Fairness concerns how errors, visibility restrictions, and opportunities to challenge decisions are distributed across users and groups. Empirical studies of warning-label systems show that labels may be applied inconsistently, including both false positives, where benign content is labelled, and false negatives, where harmful or misleading content remains unlabelled. For example, an analysis of COVID-19 warning labels on TikTok found that 37.3\% of the manually examined videos contained benign information despite receiving a warning label, while 7.7\% contained harmful or misleading information without being labelled \cite{ling2023learn}. Research on Twitter has similarly documented cases in which warning labels were incorrectly applied or omitted, including inconsistencies across languages and content formats \cite{zannettou2021won}. In the context of synthetic media, such errors may result not only in inaccurate information being shown to recipients, but also in unjustified demotion, reduced visibility, or removal of authentic content.
The distribution of these errors may be uneven. Warning-label research indicates that the effects of labels depend on their design, source, visibility, content, and the characteristics of the users who encounter them \cite{martel2023misinformation}. Political congruence and partisanship may affect how users respond to automated labels, and different label formulations can influence which groups engage with labelled content and how they interpret it \cite{papakyriakopoulos2022impact,lim2023effects}. Moreover, moderation classifications are not value-neutral. Research on sexism and misogyny classification demonstrates that annotators' attitudes affect the labels they assign and that automated models may reproduce these perspectival differences and struggle with more complex or imbalanced label sets \cite{jiang2024reexamining}. Applied to synthetic-media detection, this suggests that apparently technical categories such as ``generated,'' ``modified,'' or ``authentic-looking'' may reflect assumptions embedded in training data, annotation practices, and platform-specific moderation policies.

Procedural fairness is therefore as important as classification accuracy. Users must be able to understand why content was labelled, what elements were identified as generated or manipulated, and what consequences followed from that classification. Research on contestability in content moderation shows that affected communities seek greater representation, clearer communication, and mechanisms that allow them to shape and challenge moderation decisions \cite{vaccaro2021contestability}. Contestability should consequently be built into synthetic-media labelling systems through accessible explanations, correction and appeal mechanisms, and opportunities to provide contextual evidence.
\section{Statement of Reasons Data and Reported Practice}
The DSA Statement of Reasons database provides an initial view of how synthetic-media moderation is reported in practice \cite{noauthor_statements_nodate}. In a filtered view for the content type ``Synthetic Media'', covering the period from 25 September 2023 to 28 April 2026, the database shows 153,138,573 statements of reasons (see Appendix~\ref{screenshots}). This indicates that synthetic media is already being captured as a distinct category in the DSA transparency infrastructure.
Facebook accounts for by far the largest share, with approximately 118.33 million statements of reasons, followed by Instagram with approximately 31.36 million. X accounts for approximately 2.74 million, and Threads for approximately 717,000. Other listed platforms, including Pornhub, Takeaway.com Central Core B.V., and Upwork, appear only marginally in the filtered view.

The restriction data further shows that synthetic-media moderation is predominantly implemented as a visibility intervention. Visibility measures account for approximately 151.46 million cases, or 98.9\% of all statements of reasons in the filtered view. Combined visibility, service-provision, and account restrictions account for approximately 838,000 cases, or 0.55\%. A more detailed breakdown shows that removal of content is the dominant measure, with approximately 129.79 million statements of reasons. Demotion of content follows with approximately 19.08 million cases. Labelled content appears much less frequently, with approximately 1.42 million statements of reasons. Other visibility restrictions account for approximately 1.16 million cases.
This distribution is significant for the present analysis. Although legal and policy debates often foreground labelling as a central response to synthetic media, the DSA database suggests that platforms more often report removal or demotion than labelling as the actual moderation response. Labelling is present, but it appears as a relatively small part of the overall synthetic-media moderation response.
The information-source data also provides useful insight into how synthetic-media moderation is triggered.

Approximately 150.10 million statements of reasons, or 98.02\%, are based on own initiative. Other notification types account for approximately 299,000 cases, or 0.20\%, while Article~16 notices and trusted flagger notices appear marginal in the visible breakdown. This suggests that synthetic-media moderation is largely platform-initiated rather than user- or trusted-flagger-initiated. For user-facing labelling mechanisms, this raises the question of how much practical work is actually done by recipient-side indication tools, as opposed to automated or internal platform detection and enforcement systems.
\paragraph{Scope of the dashboard evidence.}
These figures describe statements submitted by platforms and classified under the dashboard's ``Synthetic Media'' content type; they do not measure the prevalence of synthetic media or establish that a restriction was imposed because content was synthetic. The counts concern statements, rather than unique items of content, and reflect platform reporting practices. Accordingly, the relatively low number of reported labelling measures cannot establish how often platforms label synthetic media outside this reporting dataset. The screenshots reproduce rounded dashboard values, and the analysis does not independently validate the underlying records.

\section{Preliminary Argument}
Deepfake labelling produces four governance tensions.
\begin{itemize}
   \item First, it creates \textbf{definitional ambiguity}: the relevant question is not merely whether AI was used, but whether it materially changes the communicative status of content. Borderline cases such as beauty filters, AI-generated captions, background extension, and face-swapping therefore require more granular categories than ``real'' and ``fake.''
   \item Second, labelling creates \textbf{interface and responsibility-design problems}. Recipient-facing reporting tools make users part of the labelling infrastructure, although they may be unable to identify AI-generated content reliably or may strategically misuse deepfake allegations. Labels are therefore interface interventions whose visibility, salience, and timing shape their effect.
   \item Third, labelling raises \textbf{questions of usefulness and communicative effect}. Its value depends on whether users notice, understand, and appropriately interpret the disclosure. In political contexts, claims that authentic content is AI-generated may also become a strategy of denial.
   \item Fourth, labelling raises \textbf{fairness and contestability concerns}. Classification errors may be unevenly distributed across languages, cultural contexts, and forms of expression, while false labels may lead to demotion, removal, or reputational harm. Fair implementation therefore requires accessible explanations and meaningful opportunities to challenge erroneous decisions.
\end{itemize}
\section{Conclusion}
EU deepfake regulation is evolving from disclosure duties into a broader regime of provenance, marking, detection, and interface design. The DSA addresses platform-level mitigation, while the AI Act and Code of Practice distinguish provider-side marking from deployer-side labelling. The key challenge is not only whether content is labelled, but who triggers the label, how it is interpreted, and how it interacts with moderation and transparency mechanisms. Meaningful implementation therefore requires clear, contestable, and user-facing disclosure systems.

\bibliographystyle{plainnat}
\bibliography{references}
\clearpage
\appendix
\newgeometry{margin=15mm}
\begin{landscape}
\section{EU Icon Categories and Regulatory Thresholds}
\begingroup
\footnotesize
\renewcommand{\arraystretch}{1.05}
\setlength{\parskip}{0pt}
\setlength{\tabcolsep}{5pt}
\begin{longtable}{@{}>{\RaggedRight\arraybackslash}p{0.11\linewidth}>{\RaggedRight\arraybackslash}p{0.205\linewidth}>{\RaggedRight\arraybackslash}p{0.195\linewidth}>{\RaggedRight\arraybackslash}p{0.215\linewidth}>{\RaggedRight\arraybackslash}p{\dimexpr0.275\linewidth-40pt\relax}@{}}
\caption{EU icon categories in relation to the AI Act, the DSA, the second draft Code of Practice, and the threshold analysis developed in this paper.}\label{tab:eu-icons-legal-framework}\\
\toprule
\textbf{Category} & \textbf{AI Act} & \textbf{DSA} & \textbf{Code of Practice} & \textbf{Conceptual dimension} \\ \midrule
\endfirsthead
\multicolumn{5}{l}{\textit{Table \thetable\ continued}}\\
\toprule
\textbf{Category} & \textbf{AI Act} & \textbf{DSA} & \textbf{Code of Practice} & \textbf{Conceptual dimension} \\ \midrule
\endhead
\midrule\multicolumn{5}{r}{\textit{Continued on next page}}\\\endfoot
\bottomrule\endlastfoot
\textbf{Basic AI icon} &
Articles~50(4) and 50(5) require deployers to disclose qualifying deepfakes and public-interest text clearly, distinguishably, accessibly, and no later than first exposure. &
Article~35(1)(k) refers to prominent interface markings and easy-to-use functionality through which recipients can indicate generated or manipulated content. &
The basic icon may accompany a customised textual disclosure or an interactive second layer. The Code specifies design and placement requirements for such disclosures
\cite[pp.~29--33, 37]{CodePracticeTransparencyAIGeneratedContent2026SecondDraft}. &
Supports context-specific disclosure where a generic label would not sufficiently explain which element was generated or modified. \\

\textbf{Fully AI-generated}\par
\textbf{``AI GENERATED''} &
Article~50(2) requires provider-side machine-readable marking and detectability. Article~50(4) adds visible disclosure where the output constitutes a deepfake or qualifying public-interest text. &
Article~35(1)(k) applies where generated image, audio, or video appreciably resembles existing persons, objects, places, entities, or events and falsely appears authentic or truthful. &
The Code provides for digitally signed metadata, imperceptible watermarking, and detection mechanisms. The Annex identifies the ``AI GENERATED'' icon for fully generated content
\cite[pp.~8--15, 37]{CodePracticeTransparencyAIGeneratedContent2026SecondDraft}. &
Represents the clearest case in which content is ``AI-generated enough.'' Remaining questions concern resemblance, false authenticity, publication purpose, and statutory exceptions. \\

\textbf{Partially AI-modified}\par
\textbf{``AI MODIFIED''} &
Articles~3(60) and 50(4) cover AI-generated or manipulated media that resembles existing persons, objects, places, entities, or events and falsely appears authentic or truthful. &
Article~35(1)(k) similarly covers materially generated or manipulated media satisfying the resemblance and false-authenticity criteria. &
The ``AI MODIFIED'' icon applies to pre-existing human-created content that has been partially modified with AI. A second layer may explain the nature and location of the modification
\cite[pp.~33, 37]{CodePracticeTransparencyAIGeneratedContent2026SecondDraft}. &
Central to the threshold problem. Disclosure should depend on whether AI materially changes a represented person, event, statement, or evidentiary claim. \\

\textbf{AI-assisted or minor enhancement} &
Article~50 does not make every use of AI subject to visible disclosure. Article~50(4) depends on whether the content constitutes a deepfake or qualifying manipulated public-interest text. &
Article~35(1)(k) is not triggered merely because AI was used; the resemblance and false-authenticity requirements must also be satisfied. &
The Code does not establish a separate icon for minor correction, noise reduction, automated subtitles, or limited enhancement, but encourages richer provenance and layered information
\cite[pp.~11, 33, 37]{CodePracticeTransparencyAIGeneratedContent2026SecondDraft}. &
Illustrates the unresolved boundary. Raises questions about how disclosure should depend on materiality, communicative significance, and the risk of false authenticity rather than mere AI involvement. \\

\textbf{Recipient indication} &
The primary duties remain with providers and deployers; recipients are not responsible for determining whether content legally qualifies as a deepfake. &
Article~35(1)(k) requires easy-to-use recipient functionality. Articles~16 and 22 may also be relevant where notices concern illegal content. &
Detection solutions should be available to users, researchers, civil society, media, fact-checkers, trusted flaggers, and authorities
\cite[pp.~12--15]{CodePracticeTransparencyAIGeneratedContent2026SecondDraft}. &
Creates interface and responsibility-design problems. Users need usable categories, explanations, detection support, and mechanisms to challenge erroneous decisions. \\

\textbf{Machine-readable provenance} &
Article~50(2) requires machine-readable marking and detectability, separately from visible deployer disclosure under Article~50(4). &
Technical provenance may support Article~35 mitigation, but invisible metadata alone does not necessarily satisfy the objective of prominent interface disclosure. &
The Code provides for digitally signed metadata, imperceptible watermarking, optional fingerprinting or logging, and richer provenance information
\cite[pp.~8--11]{CodePracticeTransparencyAIGeneratedContent2026SecondDraft}. &
Shows that provenance and visible labelling perform different functions. Technical origin information does not by itself determine whether content is misleading, harmful, satirical, or legally subject to disclosure. \\

\end{longtable}
\endgroup
\end{landscape}
\restoregeometry
\clearpage
\section{Statement of Reasons Dashboard Screenshots}
\label{screenshots}
All screenshots were collected on 30 April 2026. Values displayed in thousands are rounded by the dashboard. The filter dates refer to submission dates.
\begin{figure}[!ht]
\centering
\includegraphics[width=\textwidth,height=0.34\textheight,keepaspectratio]{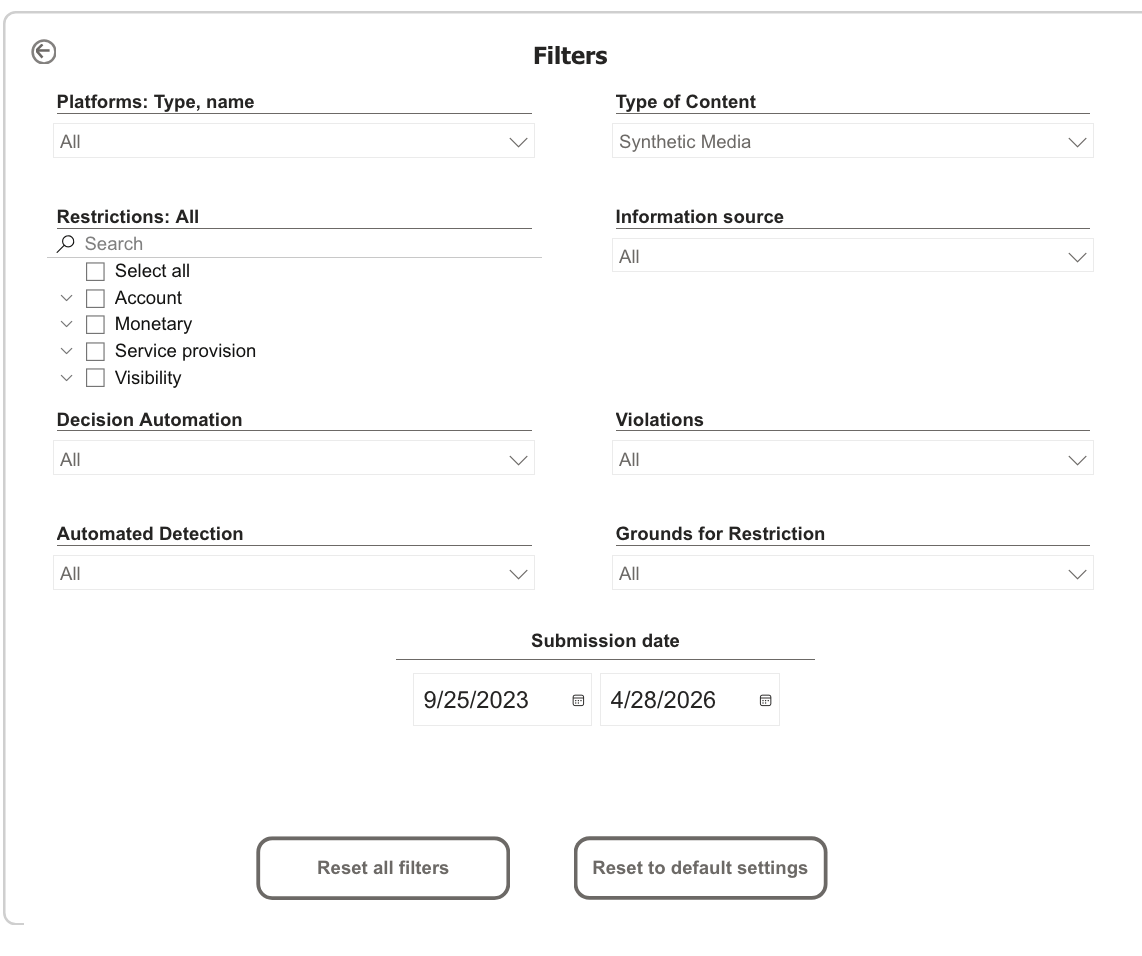}
\caption{Filter settings: content type ``Synthetic Media; submission dates 25 September 2023--28 April 2026; all platforms, restrictions, information sources, violations, grounds, and automation categories.}
\label{fig:sor-filters}
\end{figure}
\begin{figure}[!ht]
\centering
\includegraphics[width=\textwidth,height=0.34\textheight,keepaspectratio]{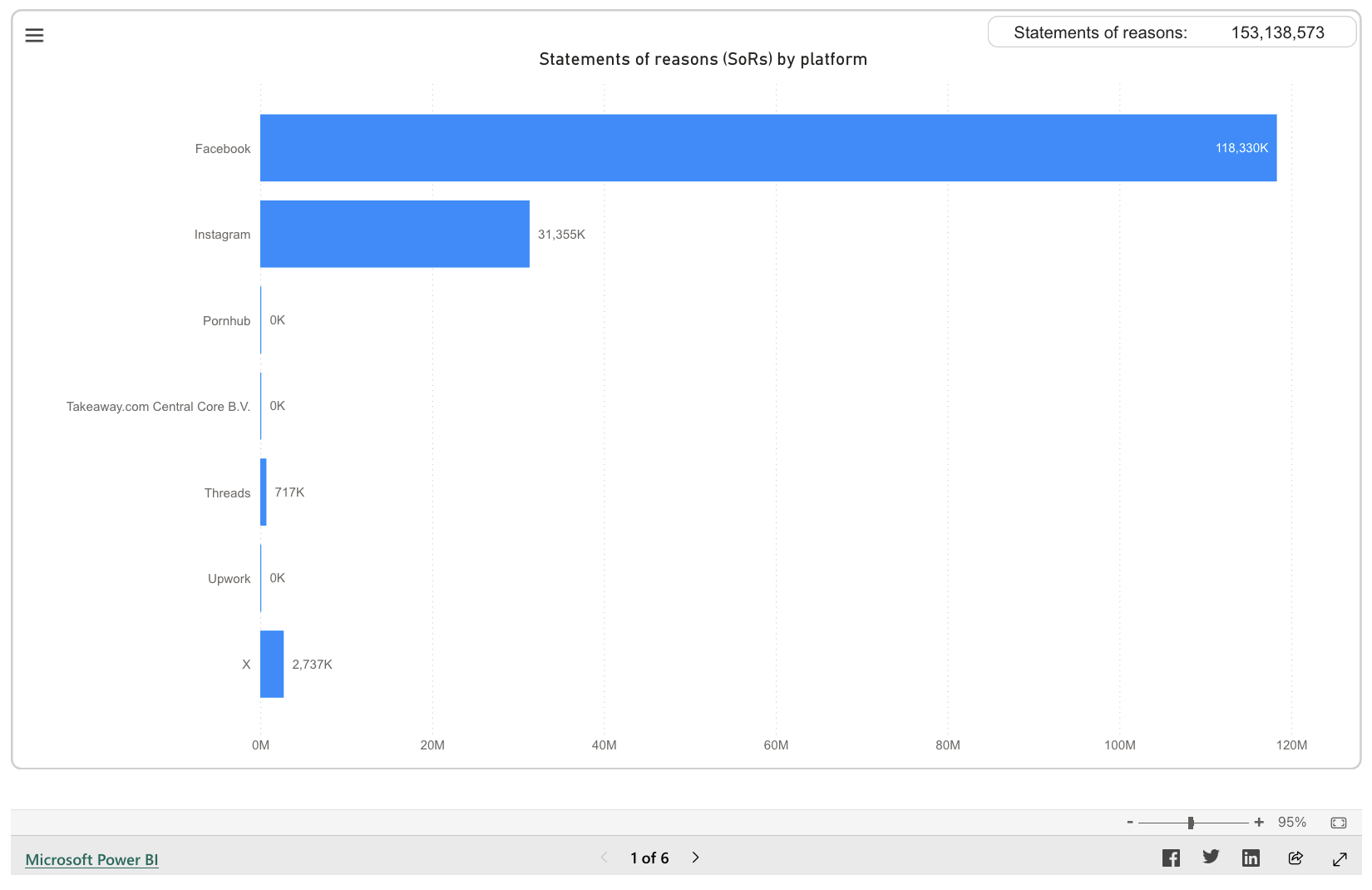}
\caption{Statements of reasons by platform. The dashboard reports 153,138,573 statements in total, concentrated on Facebook, Instagram, X, and Threads.}
\label{fig:sor-platforms}
\end{figure}
\clearpage
\begin{figure}[!ht]
\centering
\includegraphics[width=\textwidth,height=0.34\textheight,keepaspectratio]{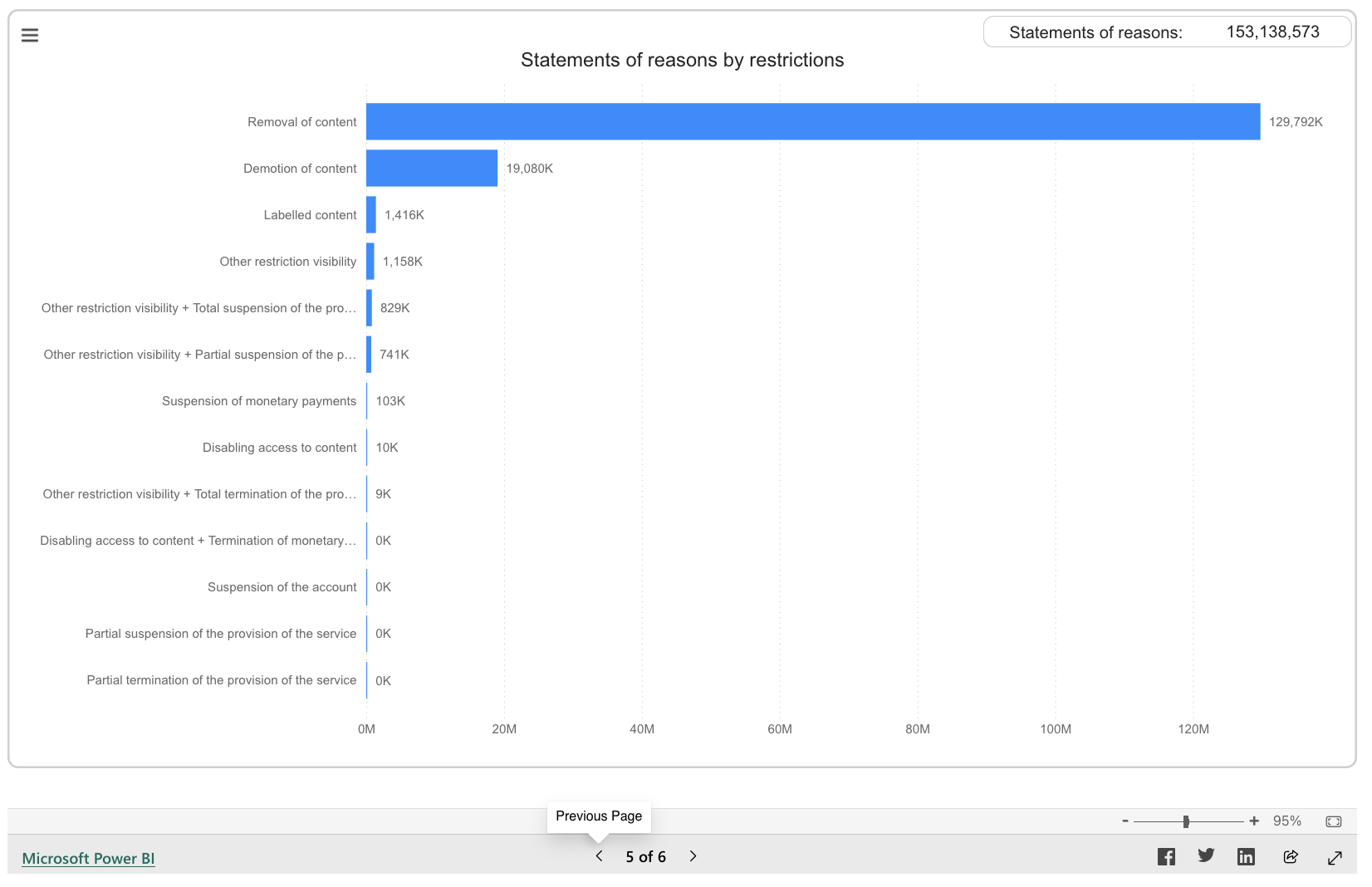}
\caption{Statements of reasons by restriction. Removal and demotion dominate; labelled content accounts for approximately 1.42 million statements.}
\label{fig:sor-restrictions}
\end{figure}
\begin{figure}[!ht]
\centering
\includegraphics[width=\textwidth,height=0.34\textheight,keepaspectratio]{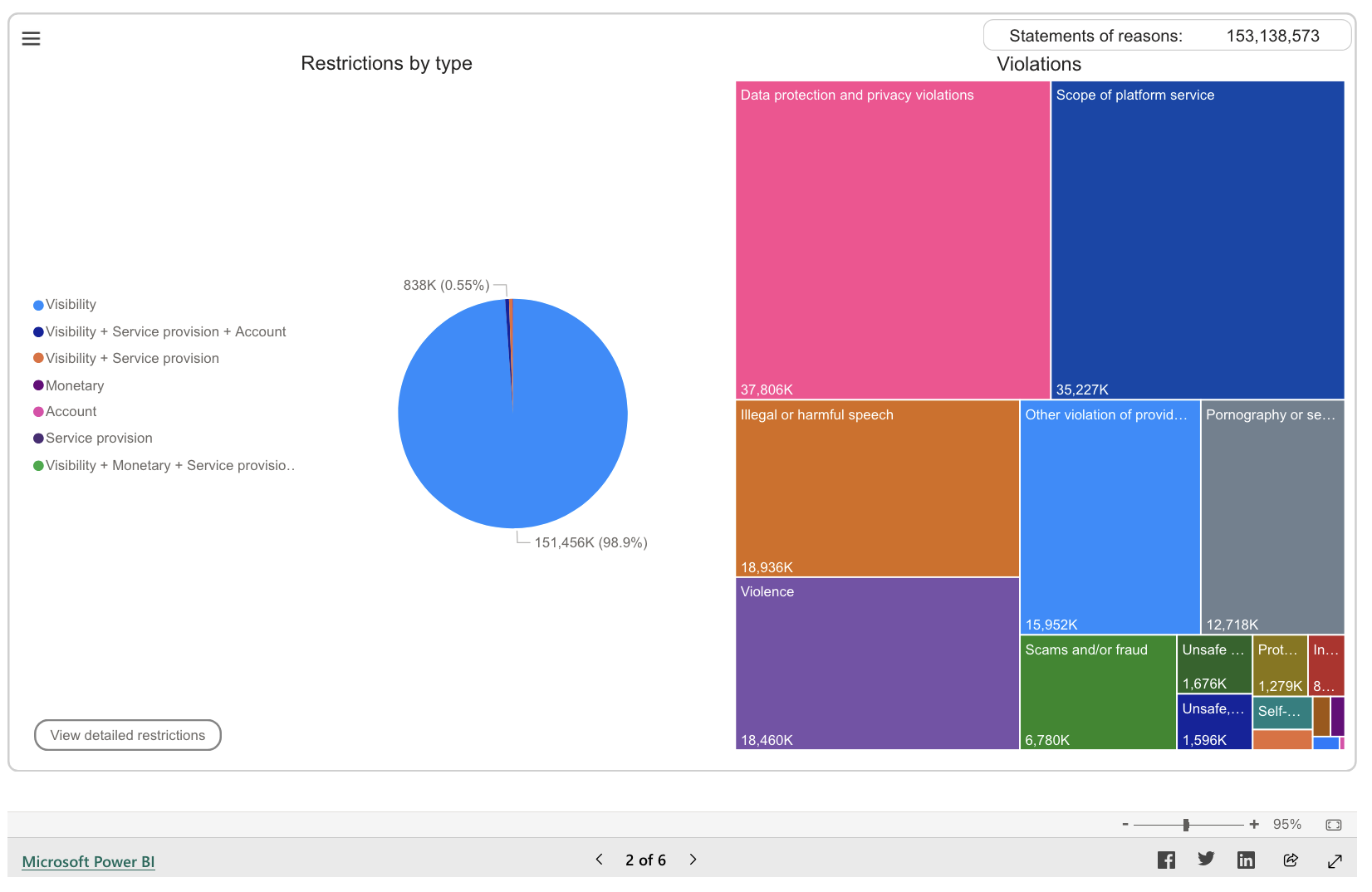}
\caption{Restriction types and violation categories. Visibility-only restrictions account for approximately 98.9\% of statements; the treemap displays the reported violation categories.}
\label{fig:sor-violations}
\end{figure}
\clearpage
\begin{figure}[!ht]
\centering
\includegraphics[width=\textwidth,height=0.34\textheight,keepaspectratio]{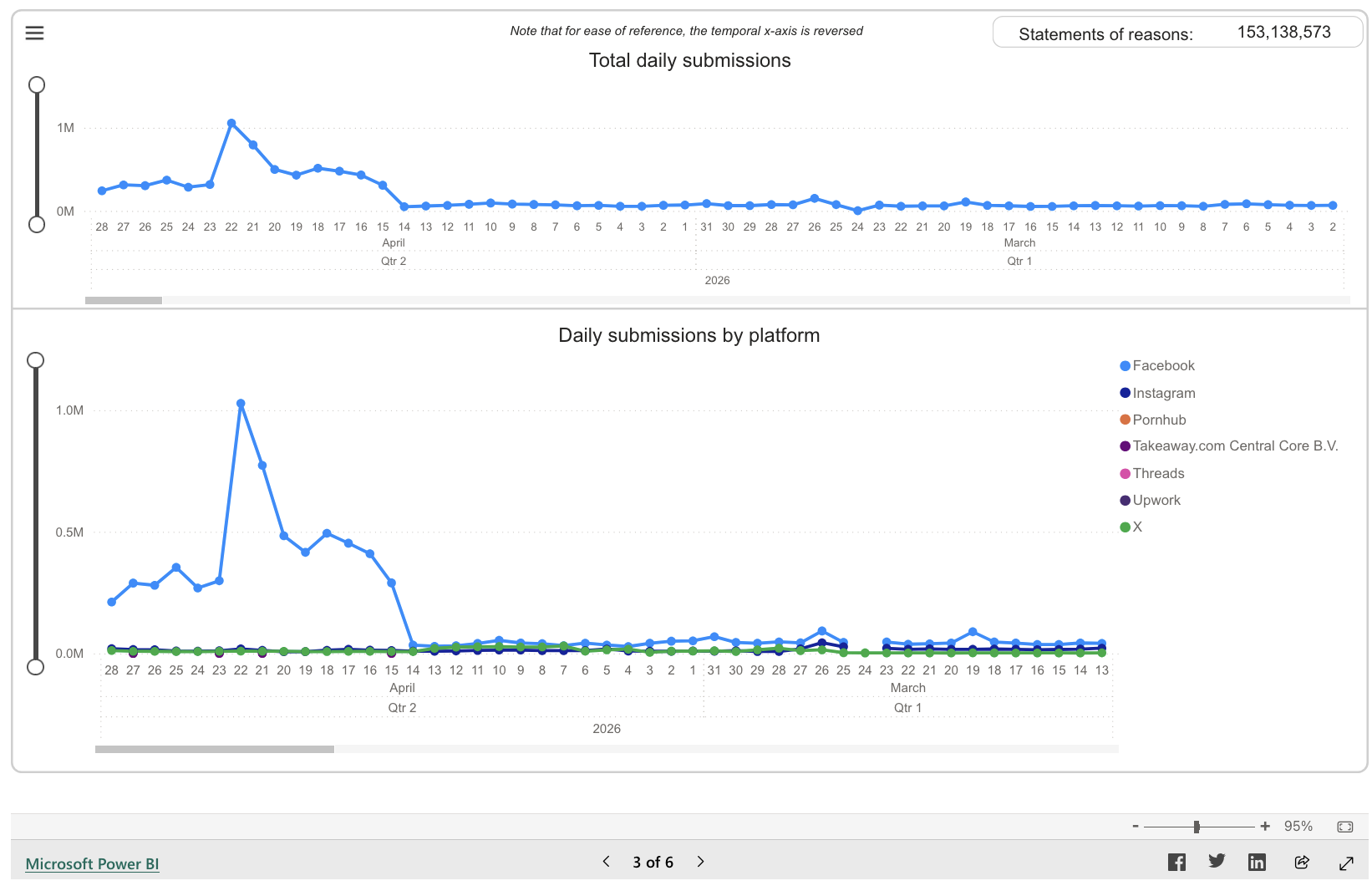}
\caption{Daily submissions, overall and by platform. The temporal axis is reversed. The displayed late-April spike is driven primarily by Facebook.}
\label{fig:sor-daily-platforms}
\end{figure}
\begin{figure}[!ht]
\centering
\includegraphics[width=\textwidth,height=0.34\textheight,keepaspectratio]{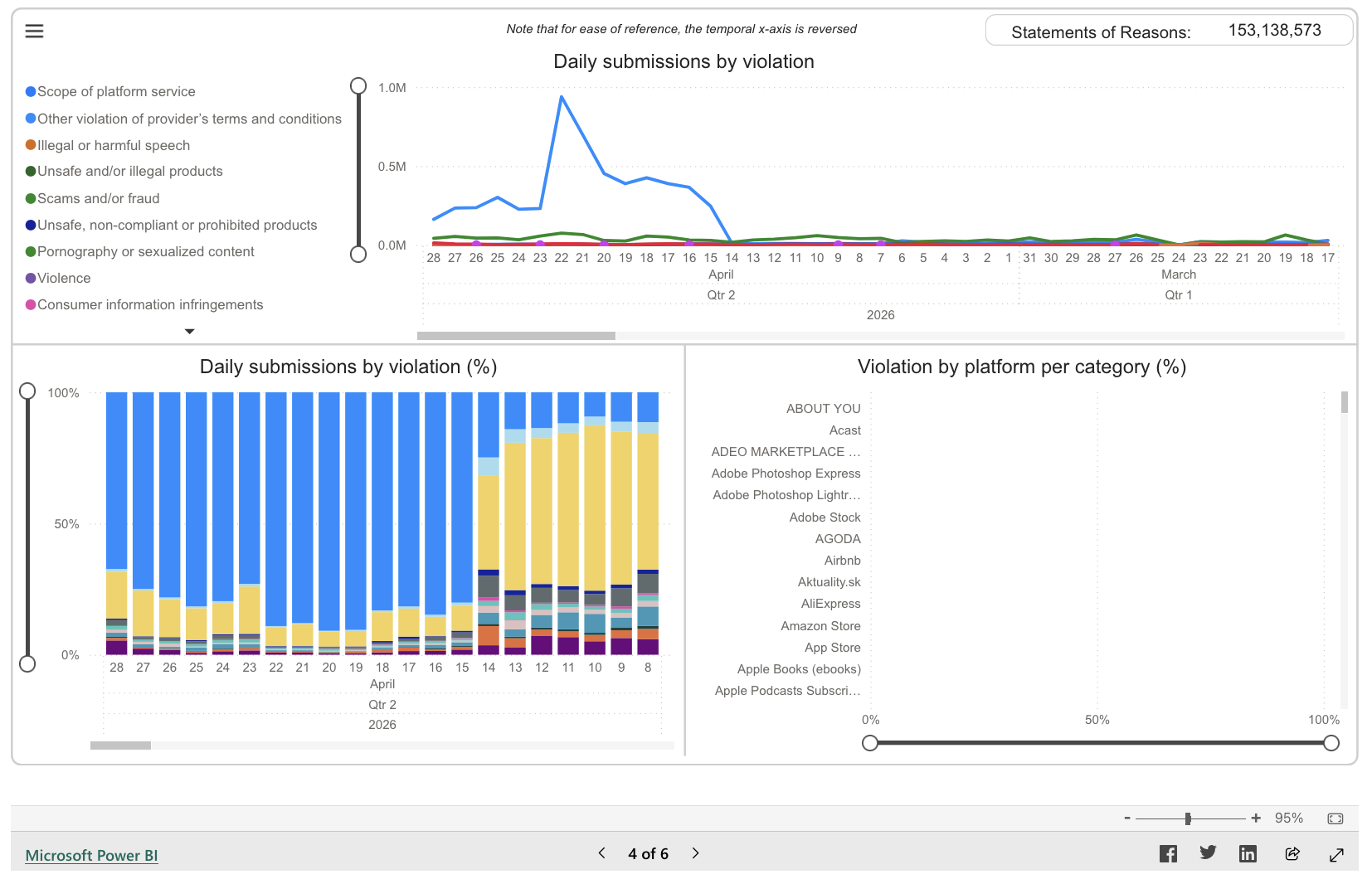}
\caption{Daily submissions by violation category, in absolute and proportional terms. The temporal axis is reversed; the lower-right panel shows violations by platform.}
\label{fig:sor-daily-violations}
\end{figure}
\clearpage
\begin{figure}[!ht]
\centering
\includegraphics[width=\textwidth,height=0.34\textheight,keepaspectratio]{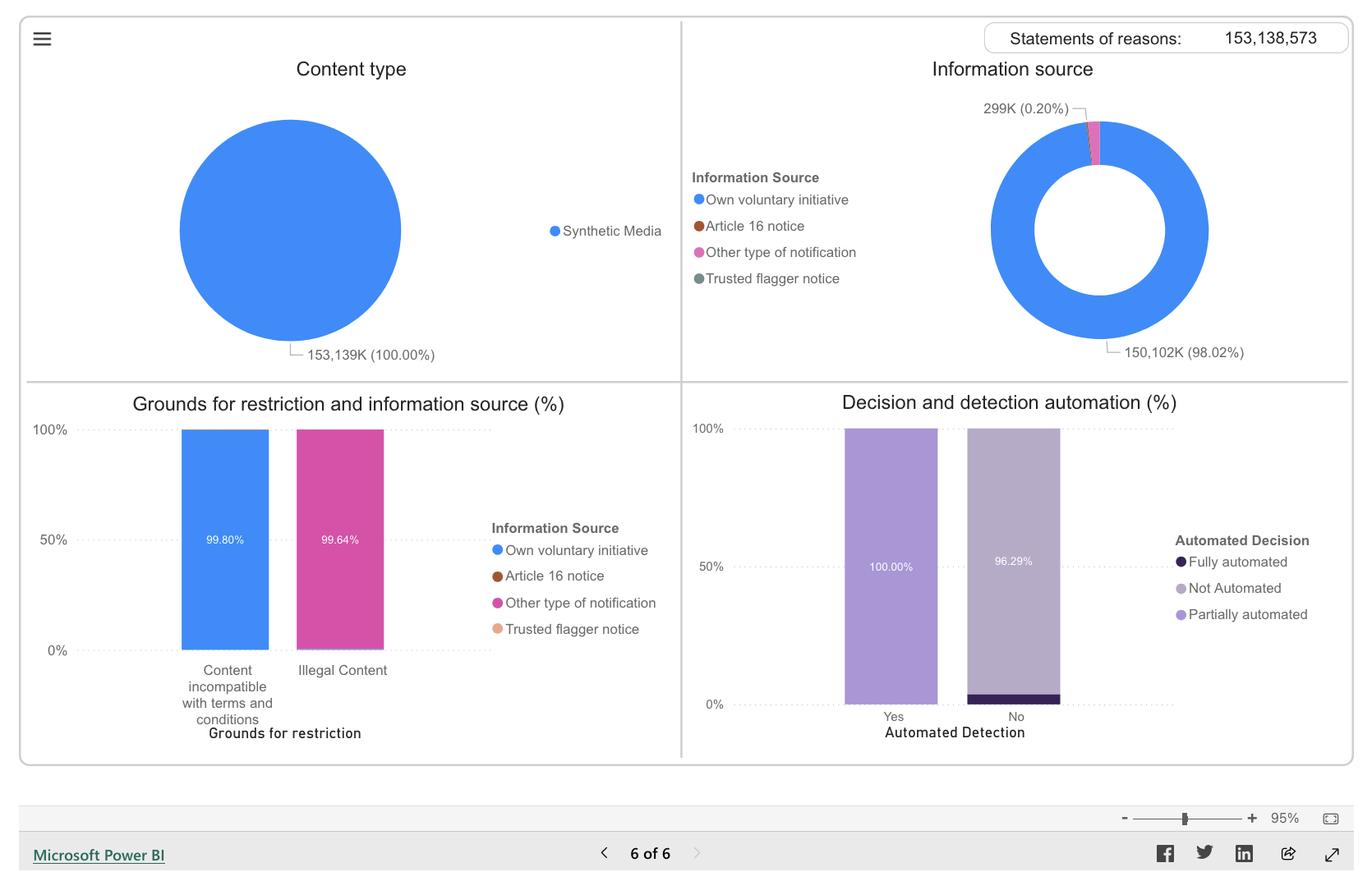}
\caption{Content type, information source, grounds for restriction, and decision and detection automation. Own voluntary initiative accounts for approximately 98.02\% of reported statements.}
\label{fig:sor-sources}
\end{figure}

\end{document}